# A Visual Language for Composable Inductive Programming

Edward McDaid FBCS
Chief Technology Officer
Zoea Ltd

Sarah McDaid PhD
Head of Digital
Zoea Ltd

**Abstract**

We present Zoea Visual which is a visual programming language based on the Zoea composable inductive programming language. Zoea Visual allows users to create software directly from a specification that resembles a set of functional test cases. Programming with Zoea Visual involves the definition of a data flow model of test case inputs, optional intermediate values, and outputs. Data elements are represented visually and can be combined to create structures of any complexity. Data flows between elements provide additional information that allows the Zoea compiler to generate larger programs in less time. This paper includes an overview of the language. The benefits of the approach and some possible future enhancements are also discussed.

## 1. Introduction

This paper describes the visual programming language Zoea Visual. In order to understand the motivation and design of Zoea Visual it is first necessary to briefly recap the underlying Zoea language.

Zoea is a simple declarative programming language that allows a user to construct software directly from a specification that resembles a set of functional test cases [1]. It is based on the concept of inductive programming [2]. Given a set of input and output examples the Zoea compiler uses AI to generate hypothesises about the required code. Zoea uses inductive programming to provide a complete, general purpose programming language. In doing this it recognises the active role of the developer in the software development process. To this end it includes the ability to specify any number of optional intermediate values between the input and the output of a test case. It also allows the developer to compose larger programs by combining any number of smaller programs. As a result Zoea is characterised as a composable inductive programming language. Composability enables Zoea to be used to construct software of any size and complexity.

The syntax of Zoea is similar to YAML with the exception that program layout is not significant [3]. A small set of tags denote the program name, test case identifiers, inputs, derived values and outputs. Data is represented using JSON [4]. Listing 1 shows a simple example of a complete Zoea program. The Zoea language has no variables, statements, conditionals, control structures or functions. Instead the developer simply provides a number of test cases showing examples of inputs and outputs as static data. The Zoea compiler uses programming knowledge, pattern matching and abductive reasoning [5] to automatically identify the required code [1]. Knowledge sources are organised as a distributed blackboard architecture [6]

The Zoea language is around 30% as complex as the simplest conventional programming languages [7]. Zoea programs are on average about the same size as equivalent programs in conventional languages but are 50% less complex. As a result Zoea is simple to learn and comparatively easy to use. Having said that the syntax of JSON represents around 80% of the grammar complexity of the Zoea language.

```
program: is_week_day
# determines if input is a weekday
  data: [ monday,tuesday,wednesday,
          thursday,friday,saturday,
          sunday ]
  case: 1  input: thursday
           output: weekday
  case: 2  input: 'MONDAY'
           output: weekday
  case: 3  input: banana
           output: unrecognised
  case: 4  input: ''
           output: unrecognised
```

**Listing 1: Example Zoea Program**

Zoea was created with the goal of making software development simpler and accessible to many more people. Learning Zoea certainly takes less time and effort than conventional programming languages. Yet most of this residual complexity involves the syntax used for data representation. In this context Zoea Visual was developed to further improve the usability of the Zoea language.

## 2. Visual Programming

Visual representations of programs have co-existed with software for most of its history. Flowcharts were widely used for specification and documentation purposes by the time that high level languages were introduced in the 1950s. Interactive creation of programs as diagrams was introduced by Sutherland in the 1960s as one of the first applications of the enabling graphical display and input technologies [8]. Visual programming became more widespread in the 1980s and 1990s with the introduction of graphical workstations and personal computers [9]. There are now many visual programming languages [10, 11,12]. Many of which are aimed at education or niche domains [13,14,15,16,17].

Visual programming languages can be characterised and evaluated with respect to a number of dimensions [18]. In this paper we will focus on the following characteristics as being the most relevant:

- Literal to abstract representation;
- Partial to complete programming language;
- Specialist domain or general purpose.

Literal languages such as those that resemble flow charts often have a clear correspondence between visual and conventional language elements. Abstract languages, such as use case diagrams, are more remote from the code they relate to and may include non-software elements such as users.

The source code of a conventional program often encodes several orthogonal aspects of software including program statements, program structure, variable scope, control flow, threading, message passing and data flow. Visual languages do not always need to describe a complete system. For example, entity relationship and class diagrams are widely used in some circles to describe database schemas and class models respectively. UML, for example, includes a number of different diagram types to model different system perspectives. While it is possible to generate fragments of code from UML diagrams it is not in itself a complete programming language.

Special purpose or domain specific visual languages have also been developed. One interesting early example was the Speech Knowledge Interface [19] which demonstrated that complex knowledge elicitation could be successfully achieved by providing the domain expert with a bespoke visual language. Many other domain specific visual languages have been developed.

Visual languages that try to be complete, general purpose programming languages face a number of challenges. Algorithmic information theory [20,21] suggests that any single notation used to describe software must either be as complex as the software it describes or else it must trade off generality. The use of multiple notations does not improve on this situation.

Whilst visual programming has always seemed like a good idea, it has constantly grappled with the following issues:

- Diagrams take up a lot of space on the screen compared to equivalent code;
- Software is complex and diagrams of software can quickly become incomprehensible;
- It often takes longer to produce a diagram than the equivalent code in a conventional language;
- Literal visual languages have all of the same concepts as equivalent conventional programming language (variables, conditions, control structures, etc.) so learning such a visual language is not significantly easier;
- Languages and frameworks often employ many classes and methods making it difficult to remember, find or discover relevant components and functions;
- There are often too many concepts to represent visually in a distinctive or recognisable way so visual languages often rely heavily on textual annotation.

These issues have never been completely addressed. This is mostly due to the intrinsic complexity of the various visual languages themselves. While visual programming plays an important role in some areas a complete and compelling visual programming language remains elusive.

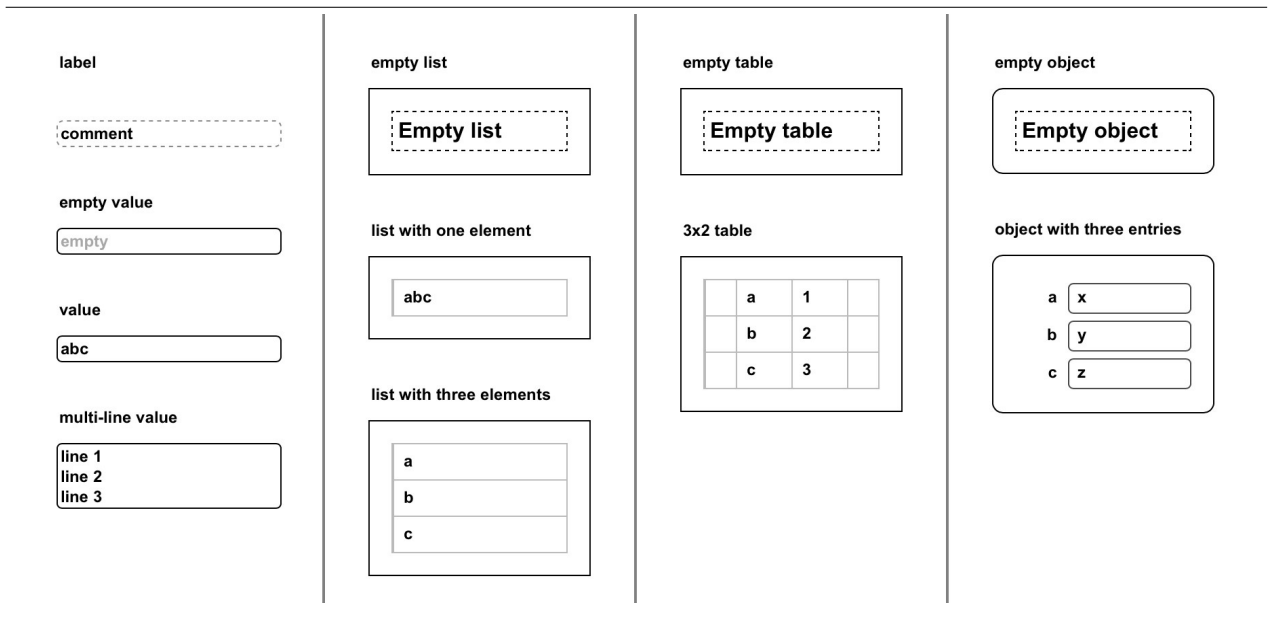

**Figure 1 – Zoea Visual Data Elements**

Abstraction is often identified as a possible way of addressing some of these problems. As with conventional programming languages there is much interest in approaches that involve different programming paradigms, higher level languages and higher order programming. Unfortunately none of these approaches have addressed the issues associated with visual programming to a significant extent.

## 3. Zoea Visual

Zoea Visual was conceived as a means of writing programs in the Zoea language using a visual programming approach. It was designed to meet the following high level requirements:

- Produce complete Zoea programs of any size;
- Submit the program to the Zoea compiler;
- Provide real-time feedback on compilation;
- Run the compiled program.

In addition it had the following non-functional requirements:

- Usable in the sense of being simple and intuitive;
- Represent programs and data visually;
- Ideally, require no change to Zoea language.

The name Zoea Visual is used to describe both the visual programming language and the associated graphical user interface.

### 3.1 Data Representation

Zoea programs consist mostly of data. The Zoea language includes a handful of tags that define the program name, case and step identifiers and program composition relationships. Everything else in a Zoea program is static data.

Data in Zoea is represented using JSON. This is a general purpose format that is also relatively simple. However large or nested data structures are not easy to comprehend even for seasoned developers. Since Zoea is also intended to be accessible for non-developers the representation of data is an important aspect of Zoea Visual.

Data elements in Zoea Visual are represented visually rather than using a serialised text format. Figure 1 provides a summary of Zoea Visual data type representations.

Individual values such as numbers and strings are represented using a textarea element. Initially this resembles a single line input field but it is configured to resize automatically to accommodate larger or multi-line values up to around 50% of the screen height. Beyond this scrolling is necessary. The user is also able to resize data elements horizontally.

One dimensional arrays (or lists) are represented as a table with a single vertical column. Each cell in the table has an input field for a single value. Similarly, two dimensional arrays are represented as a table with one or more columns. The rows represent the first dimension and the columns represent the second dimension. Additional dimensions can be supported if required by combining the appropriate number of one and two dimensional arrays. Tables have a small box at the start and end of each row. This is used to facilitate row selection. Visually it also differentiates a one dimensional array from a table with one column.

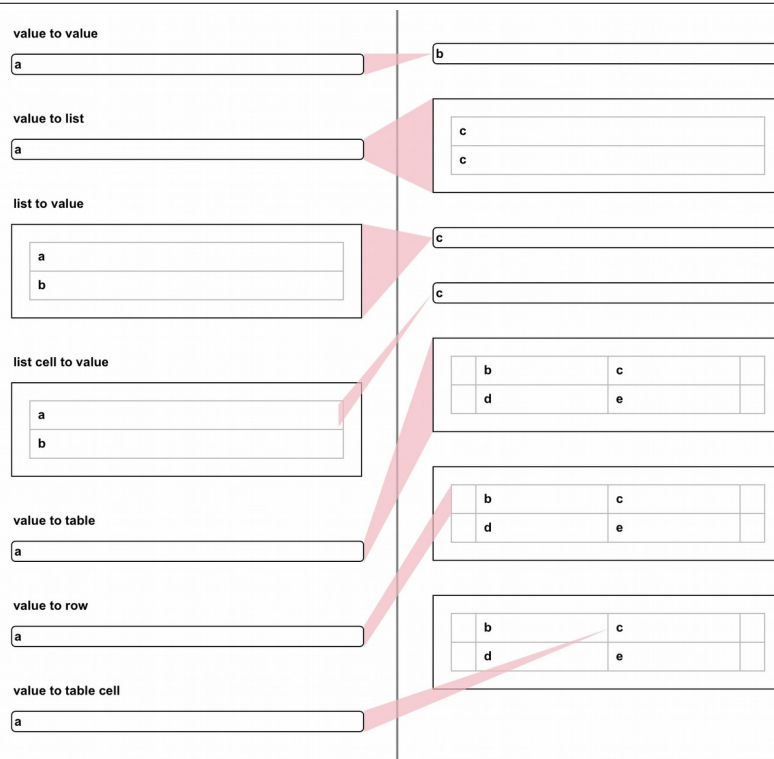

**Figure 2 – Example Zoea Visual Data Element Dependencies**

Objects (or maps) consist of a list of key-value pairs. Both the key and the value can be edited by the user. The value part is represented in the same way as a single value element while the key resembles a label.

Empty values in input fields are indicated using placeholder text to display the string 'empty' which is also coloured grey. Note that this is not an actual value so it will not be confused with a real string with the same value by the compiler. Empty lists, tables and objects also have empty element variants.

**3.2 Dependencies**

The Zoea Visual language is largely a visual representation of the Zoea language. The key difference is the additional ability in Zoea Visual to create data flow relationships between data elements. In the remainder of this paper we will refer to these links as dependencies. Figure 2 provides some examples of how dependencies between different data elements are represented in Zoea Visual.

Dependencies were added to Zoea Visual to allow the user to be more explicit about their intention with respect to the program that they are describing [22]. This has two key benefits. Firstly this additional information about the relationships between data elements allows the compiler to produce larger programs in less time. This is because dependencies allow Zoea to focus on fruitful solution candidates that utilise the correct rather than spurious data elements. The secondary role of dependencies is to make the developer intent more obvious to human readers. Greater clarity helps both the original developer and others to understand the program more easily during development and later.

Using Zoea Visual to create a program is an interactive process in which the developer is actively trying to communicate an understanding of desired program to the Zoea compiler. This is mainly achieved through the creation of test cases. Dependencies augment this process. As a Zoea developer constructs a test case they will have their own understanding of which source elements are involved in the production of a particular derived value or output. Dependencies allow the developer to capture this with little additional effort. The use of dependencies in Zoea Visual is entirely optional and the Zoea compiler will still work without them although compilation will often require more time.

However, the benefits of dependencies are such that they are likely to be used habitually.

Dependencies are represented as coloured polygons between data elements. The left side of a dependency covers the complete right side of the source element. If the target is a single value then the right side of the dependency is a point so the dependency will resemble a triangle. Dependencies with composite targets such as tables and lists instead cover the complete left side of the target and so are trapezoid in shape. This is intended to emphasise that the product is a complete composite rather than a single value.

### 3.3 Interaction

All data elements can be selected in which case they are highlighted with a yellow background. Selection is used to create dependencies or to manipulate the component in some way. Any number of data elements can be selected at a single time. Composite data elements can be selected either as a whole or in part. For example it is possible to select an entire table, or any number of rows in the table or any number of cells in any row. If a complete table is selected then any rows or cells in that table that are already selected will become deselected. Similarly if an entire row is selected then any cells in that row will be deselected. It is possible to select one or more rows in a table as well as one or more cells in different rows at the same time.

The size of composite data elements can be specified before they are added to the diagram. It is also possible to resize composite elements at any time.

Different data elements can be combined in a hierarchy to form any data structure that can be represented by JSON. For example it is possible to create a list of objects or to have a table as a value in an object, and so on. Composition is achieved by replacing the default input field in a composite data element with whichever data structure is required.

### 3.4 Inputs and Outputs

Another apparent change between the Zoea base language and Zoea Visual is that Zoea Visual supports any number of input and output values. This is accomplished by always mapping the input and output elements to lists. Zoea programs that require multiple inputs or outputs are constructed in the same way but in Zoea Visual this convention is always enforced. While this is largely a cosmetic change it goes a long way towards simplifying the user experience and is more natural for most users.

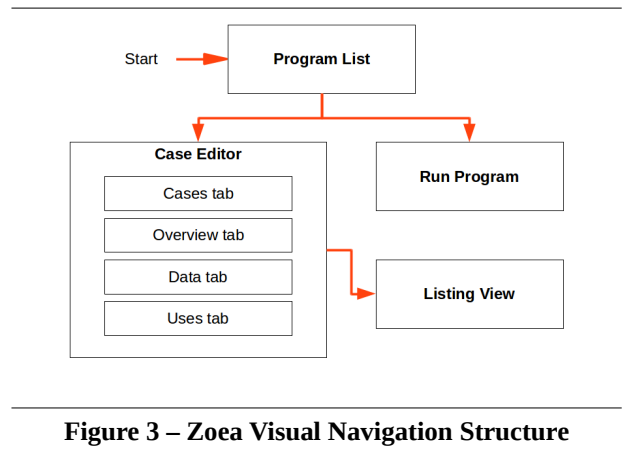

Figure 3 – Zoea Visual Navigation Structure

### 3.5 Navigation Structure

The high level navigation structure of Zoea Visual is shown in Figure 3. The entry point for users is the program screen which displays a list of existing programs. This screen also provides an option to create a new program. The case editor supports the creation of test case data models and program compilation. Once a program has been compiled it can be executed on the run screen. Other elements of the navigation structure are described later.

### 3.6 Case Editor

Each test case in a Zoea Visual program is represented by a separate diagram. The diagram is structured as a series of columns within which data values are placed. The columns correspond to the steps that are found in a Zoea program. There is a data column for constants and static reference data. This is followed by a single input column and any number of derive columns. The last column is output. The data column and any derive columns that are empty can be hidden to simplify the diagram. Figure 4 shows the Zoea Visual case editor together with a test case for a simple program.

Elements are added to a column by first selecting the appropriate column and then selecting the required component from a menu. Only a single column can be selected at a given time. If a column is selected then all data elements are deselected. Elements are removed from a column by selecting them and choosing the delete button. It is also possible to reorder elements vertically within a column. The widths of all columns can be increased or decreased to accommodate larger data structures or to make the diagram more completely visible.

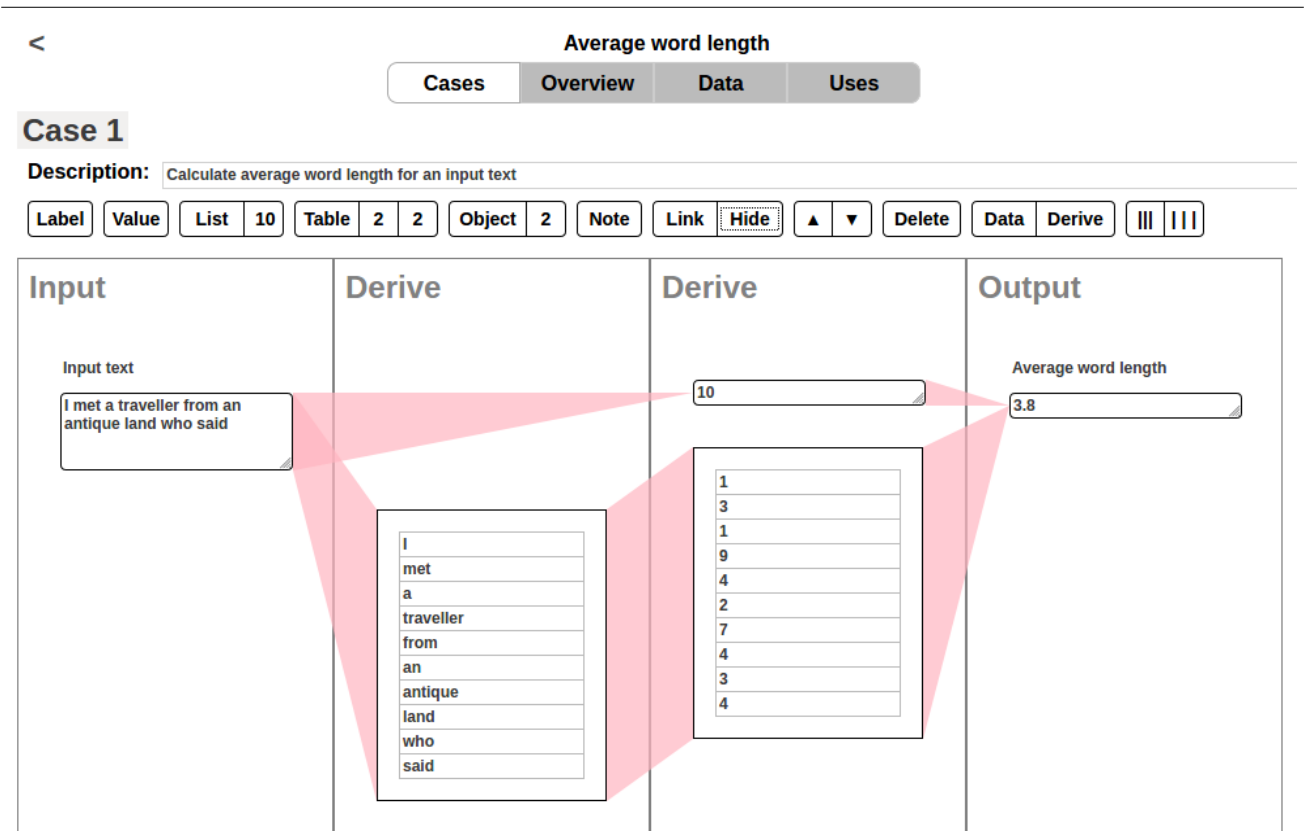

**Figure 4 – Zoea Visual Case Editor**

Dependencies are created by first selecting all of the source elements and the target element, and then selecting the link button. Element selection can be done in any order as dependencies always go from left to right across the diagram so the target element can easily be identified. It is possible to create any number of dependencies from multiple sources to a single target at the same time.

In order to help avoid dependencies covering data elements or each other it is possible to lay out the diagram to some extent. This is accomplished by moving one or more complete columns down relative to one another by varying amounts. Layout can also be adjusted by moving columns up.

Dependencies are always somewhat transparent. This allows the user to more easily discern different dependencies when they cross or when multiple dependencies originate from the same source element.

Given that dependencies from or to table cells will often cross other elements dependencies can also be temporarily faded to make them more transparent. The user can restore the appearance of dependencies to be less transparent at any time. When an element is selected any dependencies in which it is a source or the target are also highlighted.

In addition to data elements Zoea Visual allows the inclusion of comments and element labels. Comments can be placed in any column and are similar to single values except that they have a dashed border. Labels are used to describe input and output data elements at runtime and have no border. Dependencies cannot be created to or from comments or labels.

Additional cases can be added by creating new diagrams from scratch or by cloning an existing diagram. Cloning produces an exact copy of the current diagram but with empty element values. This saves considerable effort in diagram creation but more importantly the data elements and dependencies of cloned diagrams initially have exactly the same identities. This allows the compiler to recognise elements that relate to the same code more quickly. Cloned diagrams can also be edited to add and remove data elements as required.

### 3.7 Overview Tab

The case editor shows a single test case diagram at a time. In order to provide simultaneous visibility of test case data Zoea Visual also provides a tabular view of data on the overview tab. Here the values for each data element are displayed in columns with rows corresponding to each test case. This allows the user to see all of the data across all test cases.

Another role of the overview is to confirm and manage the identity of data elements across test cases. When a test case is cloned the data elements in both test cases share the same identity. If instead a new test case is created from scratch then any new data elements will have unique new identities. The overview allows the user to indicate that two or more data elements in different test cases should be treated as having the same identity. Conversely it also allows two or more elements that currently have the same identity to be treated as being distinct.

### 3.8 Runtime Data

The static data elements included in the data column are sometimes artificial and specific to the test cases that are used to specify the program. For example a Zoea program for a spell checker may use a toy dictionary with very few words as test data. This makes the program easier to specify and allows compilation to proceed more quickly. At runtime a complete dictionary can be substituted.

The data column in the case editor always displays the test version of the data. The data tab displays both the test data elements and their runtime equivalents side by side in a table. These can be edited and resized if necessary. During compilation Zoea Visual uses the test version of the data. When a compiled program is executed the runtime version of data is substituted if it has been specified.

### 3.9 Uses Tab

Zoea programs are able to import other Zoea programs with the 'use' tag. In Zoea Visual this is supported by displaying a list of compiled Zoea programs together with checkboxes. The user is able to select any number of programs to be imported into the current program.

### 3.10 Listing View

The case editor also provides a listing view in which the diagrams for all cases are displayed on a single page. Diagrams on the listing view are resized to fit the browser window. This facilitates printing and saving the diagrams as images. Dependencies on the listing view are coloured blue as they are not intended to convey information about the compilation state.

### 3.11 Compilation

The user can compile a Zoea Visual program at any time. As compilation proceeds feedback on the current status is provided to the user by colour coding the data elements and dependencies. During editing the background colour of data elements is white unless they are selected. The normal colour of dependencies is red. At the start of compilation all data elements are coloured red. The diagram for the first test case is also displayed. While a particular data element is being compiled its colour and that of any dependencies for which it is a target becomes amber. The colours of data elements and dependencies changes to green if code was successfully produced for that target element otherwise it reverts to red. To the user compilation progress is seen as moving from left to right across the diagrams. As each unique element is processed the relevant first diagram on which that element appears is also displayed.

At the end of compilation a message is displayed to inform the user of success or failure. The user will also be able to see if any of the data elements or dependencies on the current diagram are red. If compilation failed the source of the problem will be obvious to the user as the left most elements or dependencies on the current diagram that are coloured red.

As with Zoea, compilation in Zoea Visual involves the behind the scenes production of a number of synthetic test cases. A synthetic test case is a machine generated test case that consists of a subset of the data extracted directly or indirectly from a user originated test case. For example one of the compilation mechanisms employed by Zoea Visual is to construct synthetic test cases for each data element that is the target of a set of dependencies. This synthetic case has the target values of the data element across all test cases as its outputs and all of the source element values as inputs. The set of synthetic test cases can be then treated as separate and simpler problems. Each separate data element in the diagram corresponds to either a program input or a code fragment. It is not strictly necessary to always enforce left to right compilation and this aspect of Zoea Visual may be modified in the future.

The data values chosen by the user for the definition of test cases are what give Zoea its expressive capability. A single step between an input or derived value and another derived value or output can correspond to a number of statements in a conventional programming language. This includes conditionals, loops, as well as operations on sets and collections. In addition composition enables a single step to correspond to a completely separate existing program which may in turn have been composed. While it is possible for Zoea to identify multiple transformations in a single step it is probably easier for the developer if the semantics can also be easily identified by a human.

**3.12 Program Execution**

Compiled Zoea Visual programs can be run by the user. At runtime the input and output data elements and any associated labels are used to create input and output forms. Composite input data elements can also be resized as required. Input data and program details are submitted to the Zoea interpreter for execution. Results are returned to the browser and displayed to the user.

**3.13 Technology**

The technology behind the Zoea language is described in [1]. The Zoea Visual user interface is implemented in HTML5, SVG and JavaScript. It uses no third party frameworks or libraries. Communication with the Zoea compiler is over HTTPS using JSON. It can operate to some extent as an offline web application but communication with Zoea is required to load, save or execute programs.

**4. Discussion**

In terms of assessing Zoea Visual as a visual programming language it is useful to revisit the issues relating to visual programming that were identified earlier. No quantitative comparative analysis of Zoea Visual has been carried out as yet but it is possible to make some qualitative assessments.

Zoea Visual programs are likely to be smaller than equivalent programs in other visual programming languages. This is because a single dependency in Zoea Visual often corresponds to multiple statements or even complete programs in conventional programming languages.

Composable inductive programming is a programming paradigm with a much higher level of abstraction than other existing paradigms. As a result much of the complexity that would be associated with a conventional programming language is not represented explicitly in a Zoea Visual program.

Given that Zoea Visual programs are likely to be smaller and less complex than equivalent programs in other visual programming languages it is likely that they would also take less time for a user to produce them. This of course depends on the level of proficiency of the user in the respective languages.

In terms of the number of concepts that a user must learn and remember Zoea Visual is significantly simpler in terms of notation than any mainstream conventional or visual language. Aside from the static representation of data structures Zoea Visual comprises a very small number of notational elements.

When using Zoea Visual a user never has to recall the name or parameters of a particular function. Instead the user simply describes the required transformation with examples of the data. This also means that the use of text in a Zoea Visual program is limited to the specification of test data.

Both Zoea and Zoea Visual enable the user to specify a program using a set of functional test cases. Every data element in a Zoea Visual program corresponds to a fragment of the solution code. Dependencies in Zoea Visual can be considered to be subsidiary test cases that are embedded in the program level test cases. Dependencies can also span multiple test cases. This occurs frequently where the code that a dependency corresponds to is required for different cases or if the code for the dependency involves some conditional logic. As we have already noted a dependency can also correspond to a completely separate Zoea program that has already been compiled and imported. It is also possible for dependencies to overlap across test cases such that each dependency might occur in multiple test cases yet the set of dependencies for each test case remains distinct.

Zoea programs are composed with the 'use' tag. This tells the Zoea compiler that one or more specified existing Zoea programs should be treated as additional instructions and that it should try to use these in the generation of the solution. Currently the use tag is defined globally for each program. One future enhancement for Zoea Visual could be to allow composition to also be specified at the data element or dependency level. The main reason why this has not been done in the first release is that the concept of explicit dependencies is new in Zoea Visual and it was deemed important to let the concept mature properly before extending it.

Dependencies only exist in Zoea Visual. It was a deliberate decision not to introduce the concept of dependencies in the Zoea language. This is partly because it is possible to produce a Zoea program that has the same semantics using existing mechanisms. More importantly it would also require the addition of some form of data element identifiers to Zoea. This would be a big change for Zoea which was deliberately designed to avoid the exposure of users to most programming concepts including variables. Another alternative would have been to use data element paths rather than identifiers but this was considered too complex to be usable. For this reason Zoea Visual can be considered to be both an extension of the original Zoea language as well as a visual representation of it.

An important enhancement that is planned for Zoea Visual is the introduction of embedded test cases for data elements. As we have already noted each data element in Zoea Visual corresponds in some way to a piece of code in the solution. If this code is complex then we must either:

- Specify it vaguely and expect compilation to take longer and perhaps fail;
- Specify it in more detail with a greater number of program level test cases;
- Produce an external program and import it.

Embedded test cases will instead allow us to embed a complete separate test case for that data element and its associated set of dependencies. The embedded test case can have its own test cases and as much detail as we care to add including static data and derived values. It can also include further embedded test cases if necessary in the form of a hierarchy. This approach avoids the exponential growth in the number of test cases that comes with increasing complexity. It is similar to the concept of linked decision tables.

Embedded test cases will require Zoea Visual to be extended with some additional notational elements. As a minimum it will be necessary to indicate visually that a data element has an embedded test case.

Visual programming has long been considered to be a promising approach for making software development easier. While the approach has demonstrated usefulness in certain areas a complete, general purpose visual programming language that is also sufficiently usable has yet to be developed. This may simply be because existing programming paradigms do not provide a sufficient level of abstraction to make such a general purpose visual programming a usable proposition. Composable inductive programming represents a new programming paradigm that also provides a very high level of abstraction compared with other approaches.

## 5. Conclusions

Zoea Visual is a simple and effective visual language for the definition of software using the composable inductive programming approach. A large part of the language relates to the visual representation of static data. Data structures of any size and complexity can be constructed from four basic data types. Zoea Visual extends the underlying Zoea language with the concept of dependencies to describe the data flow relationships between data elements. This enables the Zoea compiler to produce larger programs in less time.

Zoea Visual represents a significant advance in terms of improving the usability of the Zoea language. Its simplicity and intuitiveness make it easy to learn and it has the potential to enable many people with no programming experience to create code.

Zoea Visual is also a significant advance in the field of visual programming languages. The combination of visual programming and composable inductive programming addresses many of the issues that have persistently dogged visual languages over the years.

## Acknowledgements

This work was supported entirely by Zoea Ltd (https://www.zoea.co.uk). Zoea is a trademark of Zoea Ltd. All other trademarks mentioned in this paper are the property of their respective owners.

The text in Figure 4 comes from Ozymandias by P.B. Shelly, (1818) The Complete Poetical Works of Percy Bysshe Shelley. Available from: http://www.gutenberg.org/ebooks/4800 (Retrieved: 15/09/2020).

## References

[1] McDaid, E., McDaid, S. (2019) Zoea – Composable Inductive Programming Without Limits. arXiv:1911.08286 [cs.PL].

[2] Kitzelmann, E. (2010) Inductive programming: A survey of program synthesis techniques. Approaches and Applications of Inductive Programming. Lecture Notes in Computer Science 5812, 50–73. Berlin, Springer-Verlag.

[3] Ben-Kiki, O., Evans, C., Döt Net, I. (2009) YAML Ain't Markup Language (YAML™)

Version 1.2. Available from: https://yaml.org/spec/1.2/spec.pdf (Retrieved: 15/09/2020)

[4] ECMA International (2017) The JSON Data Interchange Syntax. ECMA-404. 2nd edition. Available from: http://www.ecma-international.org/publications/files/ECMA-ST/ECMA-404.pdf (Retrieved: 15/09/2020)

[5] Gabriele, P. (1993) Approaches to abductive reasoning: an overview. Artificial Intelligence Review 7(2), 109-152.

[6] Nii H.P. (1986) The Blackboard Model of Problem Solving and the Evolution of Blackboard Architectures. AI Magazine 7(2), 38-53.

[7] McDaid, E., McDaid, S. (2019) Quantifying the Impact on Software Complexity of Composable Inductive Programming using Zoea. arXiv:2005.08211 [cs.PL].

[8] Sutherland, W.R. (1966) The On-line Graphical Specification Of Computer Procedures. Massachusetts Institute of Technology. Dept. of Electrical Engineering. Thesis Ph.D.

[9] Myers, B.A. (1990) Taxonomies of visual programming and program visualization. Journal of Visual Languages & Computing 1 (1), 97-123.

[10] Johnston, W.M., Hanna, J.R.P., Millar, R.J. (2004) Advances in dataflow programming languages. ACM Computing Surveys 36(1), 1–34.

[11] Patton, E.W., Tissenbaum, M., Harunani, F. (2019) MIT App Inventor: Objectives, Design, and Development. Computational Thinking Education, 31-49. Singapore, Springer.

[12] Maloney, J., Resnick, M., Rusk, N., Silverman, B., Eastmond, E. (2010) The Scratch Programming Language and Environment. ACM Transactions on Computing Education 10(4), 16, 1-15.

[13] Bockermann, C. (2014) A Visual Programming Approach to Big Data Analytics. Design, User Experience, and Usability. Lecture Notes in Computer Science 8518, 393-404. Berlin, Springer.

[14] Ray, P.P. (2017) A Survey on Visual Programming Languages in Internet of Things. Scientific Programming 2017, 1231430, 1-6. London, Hindawi. ISSN: 1058-9244.

[15] Andrade, A. (2015) Game engines: a survey. EAI Endorsed Transactions on Game-Based Learning 2(6), 150615. DOI: 10.4108/eai.5-11-2015.150615.

[16] Jeffrey., T., Kring, J. (2006) LabVIEW for everyone: graphical programming made easy and fun. (3rd ed.). New Jersey: Prentice Hall. ISBN 0131856723.

[17] Galwani, S., Hernandez-Orallo, J., Kitzelmann, E., Muggleton, S.H., Schmid, U., Zorn, B. (2015). Inductive Programming Meets the Real World. Communications of the ACM 58 (11), 90–99.

[18] Modugno, F., Green, T.R.G., Myers, B.A. (1994) Visual Programming in a Visual Domain: A Case Study of Cognitive Dimensions. People and Computers IX, Proceedings of HCI '94, 91-108. Cambridge, Cambridge University Press.

[19] Edmonds, E.A., O'Brien, S.M., Bayley, T., McDaid, E (1993) Constructing end-user knowledge manipulation systems. International Journal of Man-Machine Studies 38(1), 51-70.

[20] Kolmogorov, A. (1998) On Tables of Random Numbers. Theoretical Computer Science 207(2), 387–395.

[21] Solomonoff, R., (1964). A Formal Theory of Inductive Inference Part I. Information and Control 7(1): 1–22.

[22] Gottschlich, J., Solar-Lezama, A., Tatbul, N., Carbin, M., Rinard, M., Barzilay, R., Amarasinghe, S., Tenenbaum, J.B., Mattson, T. (2018) The Three Pillars of Machine Programming. arXiv:1803.07244 [cs.AI].